\documentclass[entropy,article,accept,moreauthors,pdftex,12pt,a4paper]{mdpi}
\lastpage{x}
\doinum{10.3390/------}
\pubvolume{xx}
\pubyear{2015}
\history{Received: xx / Accepted: xx / Published: xx}
\pdfoutput=1

\usepackage{graphicx}
\usepackage{amssymb}
\usepackage{adjustbox}
\usepackage[table]{xcolor}
\usepackage{booktabs}
\usepackage{color}

 \theoremstyle{mdpi}
 \newcounter{thm}
 \newcounter{ex}
 \newcounter{re}
 \theoremstyle{mdpidefinition}

\Title{Measuring the Complexity of Continuous Distributions}

\Author{Guillermo Santamar\'ia-Bonfil $^{1,2}$*, Nelson Fern\'andez $^{3,4}$* and Carlos Gershenson $^{1,2,5,6,7}$*}

\address{%
$^{1}$Instituto de Investigaciones en Matem\'aticas Aplicadas y en Sistemas, \\
Universidad Nacional Aut\'onoma de M\'exico.\\
$^{2}$ Centro de Ciencias de la Complejidad, UNAM, M\'exico. \\
$^{3}$ Laboratorio de Hidroinform\'{a}tica, Universidad de Pamplona, Colombia.\\
$^{4}$ Grupo de Investigaci\'on en Ecolog\'ia y Biogeograf\'ia, Universidad de Pamplona, Colombia.\\
$^{5}$ SENSEable City Lab, Massachusetts Institute of Technology, USA.\\
$^{6}$ MoBS Lab, Northeastern University, USA.\\
$^{7}$ ITMO University, St. Petersburg, Russian Federation.\\}

\corres[2]{guillermo.santamaria@iimas.unam.mx, nfernandez@unipamplona.edu.co,cgg@unam.mx.
}

\abstract{We extend previously proposed measures of complexity, emergence, and
self-organization to continuous distributions using differential entropy.
This allows us to calculate the complexity of phenomena for which
distributions are known. We find that a broad range of common parameters found in Gaussian and scale-free distributions present high complexity values.
We also explore the relationship between our measure of complexity and
information adaptation.}

\keyword{complexity; emergence; self-organization; information; differential entropy; probability distributions.}

\PACS{}

\begin{document}


\section{Introduction}

We all agree that complexity is everywhere. Yet, there is no agreed
definition of complexity. Perhaps complexity is so general that it
resists definition~\cite{Cx5Q}. Still, it is useful to have formal
measures of complexity to study and compare different phenomena~\cite{Prokopenko:2008}.
We have proposed measures of emergence, self-organization, and complexity~\cite{GershensonFernandez:2012,Fernandez2013Information-Mea}
based on information theory~\cite{Shannon1948}. Shannon information
can be seen as a measure of novelty, so we use it as a measure of
emergence, which is correlated with chaotic dynamics. Self-organization
can be seen as a measure of order~\cite{GershensonHeylighen2003a},
which can be estimated with the inverse of Shannon's information and
is correlated with regularity. Complexity can be seen as a balance
between order and chaos~\cite{Langton1990,Kauffman1993}, between
emergence and self-organization~\cite{LopezRuiz:1995,Fernandez2013Information-Mea}.

We have studied the complexity of different phenomena for different
purposes~\cite{Zubillaga2014Measuring-the-C,Amoretti2015Measuring-the-c,Febres2013Complexity-meas,Santamaria-Bonfil2016Wind-speed-fore,FernandezCxLakes}. Instead of searching for more
data and measure its complexity, we decided to explore different distributions
with our measures. This would allow us to study broad classes of dynamical
systems in a general way, obtaining a deeper understanding of the
nature of complexity, emergence, and self-organization. Nevertheless,
our previously proposed measures use discrete Shannon information. Even when any distribution can be discretized, this always
comes with caveats ~\cite{Cover2005}. For this reason, we base ourselves
on differential entropy ~\cite{Cover2005,Michalowicz2013} to propose measures for continuous distributions.

The next section provides background concepts related to information and entropies. Next, discrete measures of emergence, self-organization, and complexity are reviewed~\cite{Fernandez2013Information-Mea}. Section 4 presents continuous versions of these measures, based on differential entropy. The probability density functions used in the experiments are described in Section 5. Section 6 presents results, which are discussed and related to information adaptation~\cite{Haken2015} in Section 7.

\section{Information Theory}

Let us have a set of possible events whose probabilities of occurrence
are $p_{1},p_{2},\ldots,p_{n}\in P\left(X\right)$. Can we measure
the uncertainty described by the probability distribution $P\left(X\right)$?
To solve this endeavor in the context of telecommunications, Shannon proposed a measure of entropy~\cite{Shannon1948}, which corresponds to Boltzmann-Gibbs entropy in thermodynamics. This measure as originally proposed by Shannon, possess a
dual meaning of both uncertainty and information, even when the latter
term was later discouraged by Shannon himself~\cite{Heylighen2001}. Moreover,
we encourage the concept of entropy as the average uncertainty given
the property of asymptotic equipartition (described later in this
section). From an information-theoretic perspective, entropy measures
the average number of binary questions required to determine the value
of $p_{i}$ . In cybernetics, it is related to variety~\cite{Ashby1956}, a
measure of the number of distinct states a system can be in. 

In general, entropy is discussed regarding a discrete probability distribution. Shannon extended this concept to the continuous domain with differential entropy. However, some of the properties of its discrete counterpart are not maintained. This has relevant implications for extending to the continuous domain the measures proposed in~\cite{GershensonFernandez:2012,Fernandez2013Information-Mea}.
Before delving into these differences, first we introduce the discrete entropy, the asymptotic equipartition property (AEP), and the properties of discrete entropy. Next, differential
entropy is described, along with its relation to discrete entropy.

\subsection{Discrete Entropy}

Let $X$ be a discrete random variable, with a probability mass function
$p\left(x\right)=Pr\left\{ X=x\right\} ,x\in X$ . The entropy $H\left(X\right)$
of a discrete random variable X is then defined by

\begin{equation}
H\left(X\right)=-\sum_{x\in X}p\left(x\right)\log_{2}p\left(x\right).\label{eq:DscrtEntropy}
\end{equation}

The logarithm base provides the entropy's unit. For instance, base
two measures entropy as bits, base ten as nats. If the base of
the logarithm is $ \beta $, we denote the entropy as $H_{\beta}\left(X\right)$. Unless otherwise stated, we will consider all logarithms to be of base
two. Note that entropy does not depend on the value of $X$, but on the probabilities of the possible values $X$ can take. Furthermore, Eq.~\ref{eq:DscrtEntropy} can
be understood as the expected value of the information of the distribution.

\subsection{Asymptotic Equipartition Property for Discrete Random Variables}

In probability, the large numbers law states that, for a sequence
of \emph{n} i.i.d. elements of a sample $X$, the average value of
the sample $\frac{1}{n}\sum_{i=1}^{n}X_{i}$ approximates the expected
value $\mathbb{E}\left(X\right)$. In this sense, the Asymptotic Equipartition
Property (AEP) establishes that $H\left(X\right)$ can be approximated
by

\[
H\left(X\right)\mbox{=}\frac{1}{n}\log_{2}\frac{1}{p\left(X_{1},\ldots,X_{n}\right)},
\]

such that $n\rightarrow\infty$, and $x_{i}\in X$ are i.i.d. (independent and identically distributed).

Therefore, discrete entropy can be written also as

\begin{equation}
H\left(X\right)=\mathbb{E}\left[\log\frac{1}{p\left(X\right)}\right],\label{eq:ExpectedAverageUncertainty}
\end{equation}

where $\mathbb{E}$ is the expected value of $P\left(X\right).$ Consequently,
Eq.~\ref{eq:ExpectedAverageUncertainty} describes the expected or
average uncertainty of probability distribution $P\left(X\right).$

A final note about entropy is that, in general, any process that makes
the probability distribution more uniform increases its entropy~\cite{Cover2005}.

\subsection{Properties of Discrete Entropy}

The following are properties of the discrete entropy function. Proofs and details can be found in texbooks~
\cite{Cover2005}.
\begin{enumerate}
\item Entropy is always non-negative, $H\left(X\right)\geq0.$ 
\item $H_{\beta}\left(X\right)=\left(\log_{\beta}a\right)H_{a}\left(X\right).$
\item $H\left(X_{1},X_{2},\ldots,X_{n}\right)\leq\sum_{i=1}^{n}H\left(X_{i}\right),$
with equality iff $X_{i}$ are i.i.d.
\item $H\left(X\right)\leq\log\left|X\right|,$ with equality iff $X$ is
distributed uniformly over $X$.
\item $H\left(X\right)$ is concave.
\end{enumerate}

\subsection{Differential Entropy}

Entropy was first formulated for discrete random variables, and was
then generalized to continuous random variables in which case it is
called \emph{differential entropy}~\cite{Michalowicz2008}. It has
been related to the shortest description length, and thus, is similar
to the entropy of a discrete random variable~\cite{Calmet2005}. The
differential entropy $H\left(X\right)$ of a continuous random variable
$X$ with a density $f\left(x\right)$ is defined as

\begin{equation}
H\left(f\right)=H\left(X\right)=-\intop_{S}f\left(x\right)\log_{2}f\left(x\right)dx,\label{eq:DifferentialEntropy}
\end{equation}

where $S$ is the support set of the random variable. It is well-known
that this integral exists iff the density function of the random variables
is Riemann-integrable~\cite{Cover2005,Michalowicz2013}. The Riemann integral is fundamental in modern calculus. Loosely speaking, is the approximation of the area under any continuous curve given by the summation of ever smaller sub-intervals (i.e. approximations), and implies a well-defined concept of limit ~\cite{Calmet2005}.
$H\left(f\right)$
can also be used to denote differential entropy, and in the rest of the article,
we shall employ this notation.

\subsection{Asymptotic Equipartition Property of Continuous Random Variables}

Given a set of i.i.d. random variables drawn from a continuous distribution
with probability density $f\left(x\right)$, its differential entropy
$H\left(f\right)$ is given by

\begin{equation}
-\frac{1}{n}\log_{2}\left(f\left(X_{1},\ldots,X_{n}\right)\right)\rightarrow\mathbb{E}\left[\log_{2}\left(f\left(X\right)\right)\right]=H\left(f\right),
\end{equation}
such that $n\rightarrow\infty$. The convergence to expectation is
a direct application of the weak law of large numbers.

\subsection{Properties of Differential Entropy}
\begin{enumerate}
\item $H\left(f\right)$ depends on the coordinates.

For different choices of coordinate systems for a given probability
distribution $P\left(X\right)$, the corresponding differential entropies
might be distinct.

\item $H\left(f\right)$ is scale variant~\cite{Cover2005,Yeung2008}.

In this sense, $H\left(af\right)=H\left(f\right)+\log_{2}\left|a\right|$,
such that $a\neq0$.

\item $H\left(f\right)$is traslational invariant~\cite{Cover2005,Yeung2008,Michalowicz2013}.

In this sense, $H\left(f+c\right)=H\left(f\right)$.

\item $-\infty\leq H\left(f\right)\leq\infty$. ~\cite{Michalowicz2013}.

The $H\left(f\right)$ of a Dirac delta probability distribution,
is considered the lowest $H\left(f\right)$bound, which corresponds
to $H\left(f\right)=-\infty$.

\item Information measures such as relative entropy and mutual information
are consistent, either in the discrete or continuous domain~\cite{Yeung2008}.
\end{enumerate}

\subsection{Differences between Discrete and Continuous Entropies}

The derivation of equation~\ref{eq:DifferentialEntropy} comes from
the assumption that its probability distribution is Riemann-integrable.
If this is the case, then differential entropy can be defined just like discrete
entropy. However, the notion of ``average uncertainty'' carried
by the Eq.~\ref{eq:DscrtEntropy} cannot be extended to its differential
equivalent. Differential entropy is rather a function of the parameters
of a distribution function, that describes \emph{how uncertainty changes}
as the parameters are modified~\cite{Cover2005}.

To understand the differences between Eqs.~\ref{eq:DscrtEntropy}
and~\ref{eq:DifferentialEntropy} we will quantize a probability density
function, and then calculate its discrete entropy~\cite{Cover2005,Michalowicz2013}.

First, consider the continuous random variable $X$ with a probability
density function $f\left(x\right).$This function is then quantized
by dividing its range into \emph{h} bins of length $\Delta$. Then,
in accordance to the Mean Value Theorem, within each $h_{i}$ bin
of size $\left[i\Delta,\left(i+1\right)\Delta\right]$, there exists
a value $x_{i}^{*}$ that satisfies

\begin{equation}
\intop_{i\Delta}^{\left(i+1\right)\Delta}f\left(x\right)dx=f\left(x_{i}^{*}\right)\Delta.
\end{equation}

Then, \emph{a quantized random variable} $X_{i}^{\Delta}$is defined
as

\begin{eqnarray}
X_{i}^{\Delta}=x_{i}^{*} &  & \text{if }i\Delta\leq X\leq\left(i+1\right)\Delta,
\end{eqnarray}

and, its probability is

\begin{equation}
p_{i}=\int_{i\Delta}^{\left(i+1\right)\Delta}X^{\Delta}=f\left(x_{i}^{*}\right)\Delta.
\end{equation}

Consequently, the discrete entropy of the quantized variable $X^{\Delta}$,
is formulated as

\begin{eqnarray}
H\left(X^{\Delta}\right) & = & -\sum_{-\infty}^{\infty}p_{i}\log_{2}p_{i}\nonumber \\
 & = & -\sum_{-\infty}^{\infty}\left(f\left(x_{i}^{*}\right)\Delta\right)\log_{2}\left(f\left(x_{i}^{*}\right)\Delta\right)\nonumber \\
 & = & -\sum\Delta f\left(x_{i}^{*}\right)\log_{2}f\left(x_{i}^{*}\right)-\sum f\left(x_{i}^{*}\right)\Delta\log_{2}\Delta\nonumber \\
 & = & -\log_{2}\Delta-\sum\Delta f\left(x_{i}^{*}\right)\log_{2}f\left(x_{i}^{*}\right).\label{eq:QuantizedDscrtEntrpy}
\end{eqnarray}

To understand the final form of Eq.~\ref{eq:QuantizedDscrtEntrpy},
notice that as the size of each bin becomes infinitesimal, $\Delta\rightarrow0$,
the left-hand term of Eq.~\ref{eq:QuantizedDscrtEntrpy} becomes $\log_{2}\left(\Delta\right)$.
This is a consequence of 

\[
\lim_{\Delta\rightarrow0}\sum_{-\infty}^{\infty}f\left(x_{i}^{*}\right)\Delta=\int_{-\infty}^{\infty}f\left(x\right)dx=1.
\]

Furthermore, as $\Delta\rightarrow0$, the right-hand side of Eq.~\ref{eq:QuantizedDscrtEntrpy}
approximates the differential entropy of \emph{X} such that

\[
\lim_{\Delta\rightarrow0}\sum_{-\infty}^{\infty}\Delta f\left(x_{i}^{*}\right)\log_{2}f\left(x_{i}^{*}\right)=\int_{-\infty}^{\infty}f\left(x\right)\log_{2}f\left(x\right)dx.
\]

Note that the left-hand side of Eq.~\ref{eq:QuantizedDscrtEntrpy},
explodes towards minus infinity such that

\[
\lim_{\Delta\rightarrow0}\log_{2}\left(\Delta\right)\approx-\infty,
\]

Therefore, the difference between $H\left(f\right)$ and $H\left(X^{\Delta}\right)$is
$H\left(f\right)-H\left(X^{\Delta}\right)=\log_{2}\left(\Delta\right)$,
which approaches to $-\infty$ as the bin size becomes infinitesimal.
Moreover, consistently with this is the fact that the differential
entropy of a discrete value is $-\infty$~\cite{Michalowicz2013}.

Lastly, in accordance to~\cite{Cover2005}, the average number of
bits required to describe a continuous variable \emph{X} with a n-bit
accuracy (quantization) is $H\left(X\right)+n\approx H\left(f\right)$
such that

\begin{equation}
H\left(X^{\Delta}\right)'=\lim_{\Delta\rightarrow0}H\left(X^{\Delta}\right)+\log_{2}\left(\Delta\right)\rightarrow H\left(f\right).\label{eq:nBitQuanDifferentialEntropy}
\end{equation}

\section{Discrete Complexity Measures}

Emergence $E$, self-organization $S$, and complexity $C$
are close relatives of Shannon's entropy. These information-based
measures, inherit most of the properties of Shannon's discrete entropy
\cite{Fernandez2013Information-Mea}, being the most valuable one that, discrete
entropy quantizes the average uncertainty of a probability distribution.
In this sense, complexity $C$ and its related measures ($E$ and $S$) are based on a quantization of the average information contained by a process described by its probability distribution.

\subsection{Emergence}

Another form of entropy, rather related to the concept of information
as uncertainty, is called emergence $E$~\cite{Fernandez2013Information-Mea}. Intuitively,
$E$ measures the ratio of uncertainty a process produces by new information that is consequence of changes in a) \emph{dynamics} or b) \emph{scale}
\cite{Fernandez2013Information-Mea}. However, its formulation is more related to the thermodynamics entropy. Thus, it is defined as

\begin{equation}
E\equiv H\left(X\right)=-K\sum_{i=1}^{N}p_{i}\log_{2}p_{i},\label{eq:DiscreteEmergence}
\end{equation}
where $p_{i}=P\left(X=x\right)$ is the probability of the element
$i$, and $K$ is a normalizing constant.

\subsection{Multiple Scales}

In thermodynamics, the Boltzmann constant \emph{K}, is employed to
normalize the entropy in accordance to the probability of each state. However, Shannon's entropy typical formulation
\cite{Cover2005,Michalowicz2013,Haken2015} neglects the usage of
\emph{K} in Eq.~\ref{eq:DiscreteEmergence} (been its only constraint
that $K>0,$~\cite{Fernandez2013Information-Mea}). Nonetheless, for emergence as
a measure of the average production of information for a given distribution,
\emph{K} plays a fundamental role. In the cybernetic definition
of \emph{variety}~\cite{Heylighen2001}, $K$ is a function
of the distinct states a system can be, \emph{i.e.} the system's alphabet size.
Formally, it is defined as

\begin{equation}
K=\frac{1}{\log_{2}\left(b\right)},\label{eq:EntropyScaleNormalization}
\end{equation}
where $b$ corresponds to the size of the alphabet of the sample or
bins of a discrete probability distribution. Furthermore, $K$ should
guarantee that $0\leq E\leq1$, therefore, $b$ should be at least equal
to the number of bins of the discrete probability distribution. 

It is also worth noting that the denominator of Eq.~\ref{eq:EntropyScaleNormalization},
$\log_{2}\left(b\right),$ is equivalent to the maximum entropy for
a continuous distribution function, the uniform distribution. Consequently,
emergence can be understood as the ratio between the entropy for given
distribution $P\left(X\right)$, and the maximum entropy for the same
alphabet size $H\left(U\right)$~\cite{SinghEntrop2013}, this is

\begin{equation}
E=\frac{H\left(P\left(X\right)\right)}{H\left(U\right)}.\label{eq:EmergenceAsRatio}
\end{equation}

\subsection{Self-Organization}

Entropy can also provide a measure of system's organization, and its
predictability~\cite{SinghEntrop2013}. In this sense, with more uncertainty less predictability is achieved, and vice-versa. Thus, an entirely random process (\emph{e.g.} uniform distribution) has the lowest organization, and a completely deterministic system one (Dirac delta distribution), has the highest. Furthermore, an extremely organized system yields no information with respect of \emph{novelty}, while, on the other hand, the more chaotic a system is, the more information is yielded~\cite{SinghEntrop2013,Fernandez2013Information-Mea}.

The metric of self-organization $S$ was proposed to measure the organization
a system has regarding its average uncertainty~\cite{Gershenson2012,Fernandez2013Information-Mea}.
$S$ is also related to the cybernetic concept of \emph{constraint},
which measures changes in due entropy restrictions on the state space
of a system~\cite{Kauffman1993}. These constraints confine the system's behavior, increasing
its predictability, and reducing the (novel) information it provides to an
observer. Consequently, the more self-organized a system is, the less
average uncertainty it has. Formally, $S$ is defined as

\begin{equation}
S=1-E=1-\left(\frac{H\left(P\left(X\right)\right)}{H\left(U\right)}\right),\label{eq:DiscreteSelfOrganization}
\end{equation}
such that $0\leq S \leq1$. It is worth noting that, $S$ is the complement
of $E$. Moreover, the maximal $S$ (\emph{i.e.} $S=1$) is only achievable
when the entropy for a given probability density function (PDF) is such that $H\left(P\left(X\right)\right)\rightarrow0$,
which corresponds to the entropy of a Dirac delta (only in the
discrete case).

\subsection{Complexity}

Complexity $C$ can be described as a \emph{balance} between order (stability),
and chaos (scale or dynamical changes)~\cite{Fernandez2013Information-Mea}. More
precisely, this function describes a system's behavior in terms of
the average uncertainty produced by its probability distribution in
relation the dynamics of a system. Thus, the
complexity measure is defined as

\begin{equation}
C=4\cdot E\cdot S,\label{eq:DiscreteComplexity}
\end{equation}
such that, $0\leq C\leq1$.

\section{Continuous Complexity Measures}
\label{sub:Enter-Differential-Complexity}

As mentioned before, discrete and differential entropies do
not share the same properties. In fact, the property of discrete entropy
as the average uncertainty in terms of probability, cannot be extended
to its continuous counterpart. As consequence, the proposed continuous
information-based measures describe how the production
of information changes respect to the probability distribution parameters.
In particular, this characteristic could be employed as a feature
selection method, where the most relevant variables are those which
have a high emergence (the most \emph{informative}).

The proposed measures are differential emergence ($E_{D}$), differential
self-organization ($S_{D}$), and differential complexity ($C_{D}$).
However, given that the interpretation and formulation (in terms of emergence) of discrete
and continuous $S$ (Eq.~\ref{eq:DiscreteSelfOrganization}) and $C$  (Eq.~\ref{eq:DiscreteComplexity}) are the same, we only provide details
on $E_{D}$. The difference between $S_{D}$, $C_{D}$ and $S, C$ is that the former are defined on $E_{D}$, while the latter on $E$. Furthermore, we make emphasis in the definition of the normalizing
constant \emph{K}, which play a significant role in constraining $E_{D}\in\left[0,1\right]$,
and consequently, $S_{D}$ and $C_{D}$ as well.

\subsection{Differential Emergence}

As for its discrete form, the emergence for continuous random variables
is defined as

\begin{equation}
E_{D}=-K\intop_{\upsilon}^{\zeta}f\left(X\right)\log_{2}f\left(X\right),\label{eq:DifferentialEmergence}
\end{equation}
where, $\left[\upsilon,\zeta\right]$ is the domain, and \emph{K}
stands for a normalizing constant related to the distribution's alphabet
size. It is worth noting that this formulation is highly related
to the view of emergence as \emph{the ratio of information production
of a probability distribution respect the maximum differential entropy
for the same range}. However, since $E_{D}$ can be negative (\emph{i.e.} entropy
of a single discrete value), we choose $E_{D}$ such that

\begin{equation}
E'_{D}=\begin{cases}
E{}_{D} & E{}_{D}>0\\
0 & \text{otherwise}.
\end{cases}.\label{eq:DifferentialEmergencePrime}
\end{equation}

$E'{}_{D}$ is rather a more convenient function than $E{}_{D}$, as $0\leq E'{}_{D} \leq 1$.
This statement is justified in the fact that the differential entropy
of a discrete value is $-\infty$~\cite{Cover2005}. In practice, differential entropy becomes negative only when the probability distribution is extremely narrow, \emph{i.e.} there is a high probability for few states.
In the context of information changes due parameters manipulation, an $E{}_{D}<0$
means that the probability distribution is becoming a Dirac delta
distribution. For notation convenience, from now on we will employ
$E_{D}$ and $E'_{D}$ interchangeably.

\subsection{Multiple Scales}

The $K$ constant expresses the relation between uncertainty of a given $P\left(X\right)$
Defined by $H(X)$, respect to the entropy of a maximum entropy over
the same domain~\cite{SinghEntrop2013}. In this setup, as the uncertainty
grows, $E'_{D}$ becomes closer to unity.

To constrain the value of $H\left(X\right)=\left[0,1\right]$ in
the discrete emergence case, it was enough to establish the distribution's
alphabet size, \emph{b} of Eq.~\ref{eq:DiscreteEmergence}, such that
$b\geq\text{\# bins}$~\cite{Fernandez2013Information-Mea}. However, for any PDF,
the number of elements between a pair of points $a$ and $b$, such
that $a\neq b$, is infinite. Moreover, as the size of each bin becomes
infinitesimal, $\Delta\rightarrow0$, the entropy for each bin becomes
$-\infty$~\cite{Cover2005}. Also, it has been stated
that \emph{b} value should be equal to the cardinality of $X$~\cite{SinghEntrop2013},
however, this applies only to discrete emergence. Therefore,
rather than a generalization, we propose an heuristic for the selection
of a proper \emph{K} in the case of differential emergence. Moreover,
we differentiate between \emph{b} for $H\left(f\right)$, and \emph{b'}
for $H\left(X^{\Delta}\right)'$.

As in the discrete case, \emph{K} is defined as Eq.~\ref{eq:EntropyScaleNormalization}.
In order to determine the proper alphabet size \emph{b}, we propose
the next algorithm:
\begin{enumerate}
\item If we know \emph{a priori} the true $P\left(X\right)$, we calculate $H\left(f\right)$,
and \emph{$b=\left|P\left(X\right)\right|$} is the cardinality within
the interval of Eq.~\ref{eq:DifferentialEmergence}. In this sense,
a large value will denote the cardinality of an ``ghost'' sample
\cite{Michalowicz2013}%
\footnote{It is \emph{ghost}, in the concrete sense that it does not exist.
Its only purpose is to provide a bound for the maximum entropy accordingly
to some large alphabet size.%
}.
\item If we do not know the true $P\left(X\right)$, or we are interested
rather in $H\left(X^{\Delta}\right)'$ where a sample of finite size
is involved, we calculate \emph{b' as}

\begin{alignat}{1}
b'=\sum_{i}ind\left(x_{i}\right) & ,
\end{alignat}
such that, the non-negative function $ind\left(\cdot\right)$ is defined
as

\begin{alignat}{1}
ind\left(x_{i}\right)=\begin{cases}
1 & \text{iff }P\left(x_{i}\right)>0\\
0 & \text{otherwise}
\end{cases}.
\end{alignat}

For instance, in the quantized version of the standard normal distribution
($N\left(0,1\right)$), only values within  $\pm3\sigma$ satisfy this
constraint despite the domain of Eq.~\ref{eq:DifferentialEmergence}.
In particular, if we employ $b=\left|X\right|$ rather than $b'$,
we \emph{compress} the $E_D$ value as it will be shown in the next
section. On the other hand, for a uniform distribution or a power-law
(such that $0<x_{min}<x$), the whole range of points satisfies this
constraint.

\end{enumerate}



\section{Probability Density Functions}

In communication and information theory, uniform (U) and normal, a.k.a. Gaussian (G) distributions
play a significant role. Both are referent to maximum entropy:
on the one hand, U has the maximum entropy within a continuous domain;
on the other hand, G has the maximum entropy for distributions with a fixed
mean ($\mu$), and a finite support set for a fixed standard deviation
($\sigma$)~\cite{Cover2005,Michalowicz2013}. Moreover, as mentioned
earlier, $H\left(f\right)$  is useful when comparing
the entropies of two distributions over some reference space~\cite{Cover2005,Sharma2006,Michalowicz2013}.
Consequently, U, but mainly G, are heavily used in the context of
telecommunications for signal processing~\cite{Michalowicz2013}.
Nevertheless, many natural and man-made phenomena can be approximated with power-law (PL) distributions. These types of distributions typically
present complex patterns that are difficult to predict, making them a relevant research topic~\cite{Dover2004}. Furthermore, power-laws have
been related to the presence of multifractal structures in certain
types of processes~\cite{Sharma2006}. Moreover, power-laws are tightly
related to self-organization and criticality theory, and have been
studied under information frameworks before (\emph{e.g.} Tsallis', and Renyi's
maximum entropy principle)~\cite{Bashkirov2000,Dover2004}.

Therefore, in this work we focus our attention to these three PDFs.
First, we provide a short description of each PDF, then, we summarize
its formulation, and the corresponding $H\left(f\right)$ in Table
\ref{tab:PDFsandDiffEntrop}. 

\subsection{Uniform Distribution.}

The simplest PDF, as its name states, establishes that for each possible
value of \emph{X}, the probability is constant over the whole support
set (defined by the range between $a$ and $b$), and 0 elsewhere.
This PDF has no parameters besides the starting and ending points
of the support set. Furthermore, this distribution appears frequently
in signal processing as white noise, and it
has the maximum entropy for continuous random variables~\cite{Michalowicz2013}.

Its PDF, and its corresponding $H\left(f\right)$ are shown in first
row of Table~\ref{tab:PDFsandDiffEntrop}. It is worth noting that,
as the cardinality of the domain of U grows, its differential entropy
increases as well. 

\subsection{Normal Distribution.}

The normal or Gaussian distribution is one of the most important probability
distribution families~\cite{Ahsanullah2014}. It is fundamental in
the central limit theorem~\cite{Michalowicz2013}, time series forecasting
models such as classical autoregressive models~\cite{Box-Jenkins2008},
modelling economic instruments~\cite{Mitzenmacher}, encryption, modelling
electronic noise~\cite{Michalowicz2013}, error analysis and statistical hypothesis testing.

Its PDF is characterized by a symmetric, bell-shaped function whose
parameters are: location (\emph{i.e.} mean $\mu$), and dispersion (\emph{i.e.}
standard deviation $\sigma^{2}$ ). The standard normal distribution
is the simplest and most used case of this family, its parameters
are $N\left(\mu=0,\,\sigma^{2}=1\right)$. A continuous random variable $x\in X$
is said to belong to a Gaussian distribution, $X\sim N\left(\mu,\,\sigma^{2}\right)$
, if its PDF $p\left(x\right)$ is given by the one described in the
second row of Table~\ref{tab:PDFsandDiffEntrop}. As is shown in the
table, the differential entropy of G only depends on the standard
deviation. Furthermore, it is well known that its differential entropy
is monotonically increasing concave in relation to $\sigma$~\cite{Ahsanullah2014}.
This is consistent with the aforementioned fact that $H\left(f\right)$
is translation-invariant. Thus, as $\sigma$ grows, so does the value
of $H\left(G\right)$, while as $\sigma\rightarrow0$ such that $0<\sigma<1$,
it becomes a Dirac delta with $H\left(f\right)\approx0$.

\subsection{Power-Law Distribution.}

Power-law distributions are commonly employed to describe multiple
phenomena (\emph{e.g.} turbulence, DNA sequences, city populations, linguistics,
cosmic rays, moon craters, biological networks, data storage in organisms,
chaotic open systems, and so on) across numerous scientific disciplines
\cite{Bashkirov2000,Mitzenmacher2001,Dover2004,Sharma2006,Clauset2009,Frigg2011,Virkar2012}.
These type of processes are known for being scale invariant, being
the typically scales ($\alpha$, see below) in nature between one and 3.5~\cite{Bashkirov2000}.
Also, the closeness of this type of PDF to chaotic systems and
fractals is such that, some fractal dimensions are called entropy
dimensions (\emph{e.g.} box-counting dimension, and Renyi entropy)~\cite{Frigg2011}.

Power-law distributions can be described by continuous and discrete
distributions. Furthermore, Power-laws in comparison with Normal distribution,
generate events of large orders of magnitude more often, and are not
well represented by a simple mean. A Power-Law density distribution
is defined as

\begin{equation}
p\left(x\right)dx=P\left(x\leq X\leq x+dx\right)=Cx^{-\alpha}dx,
\end{equation}
such that, \emph{C} is a normalization factor, $\text{\ensuremath{\alpha}}$
is the scale exponent, and \emph{$X\mid x>x_{min}>0$} is the observed
continuos random variable. This PDF diverges as $x\rightarrow0$ ,
and do not hold for all $x\ge0$~\cite{Virkar2012}. Thus, $x_{min}$
corresponds to lower bound of a power-law. Consequently, in Table
\ref{tab:PDFsandDiffEntrop} we provide the PDF of a Power-Law as
proposed by~\cite{Clauset2009}, and its corresponding $H\left(f\right)$
as proposed by~\cite{Yapage2014}.

The aforementioned PDFs, and their corresponding $H\left(f\right)$
are shown in Table~\ref{tab:PDFsandDiffEntrop}. Further details about
the derivation of $H\left(f\right)$ for U, and G can be found in
\cite{Cover2005,Michalowicz2013}. For additional details on the differential
entropy of the power-law, we refer the reader to~\cite{Sharma2006,Yapage2014}.

\begin{table}[H]
\begin{adjustbox}{width=1\textwidth,center}

\begin{tabular}{|c|c|c|}
\hline 
Distribution & PDF & Differential Entropy\tabularnewline
\hline 
\hline 
Uniform  & $p\mbox{\ensuremath{\left(x\right)}=}\begin{cases}
\frac{1}{b-a} & a\leq x\leq b\\
0 & \text{otherwise}
\end{cases}$ & $H\left(p\left(x\right)\right)=\log_{2}\left(b\right)$\tabularnewline
\hline 
Normal  & $p\left(x\right)=\frac{1}{\sigma\sqrt{2\pi}}e^{\frac{-\left(x-\mu\right)^{2}}{2\sigma}}$ & $H\left(p\left(x\right)\right)=\frac{1}{2}\log_{2}\left(2\pi e\sigma^{2}\right)$\tabularnewline
\hline 
Power-law  & $p\left(x\right)=\left(\frac{\alpha-1}{x_{min}}\right)\left(\frac{x}{x_{min}}\right)^{-\alpha}$ & $H\left(p\left(x\right)\right)=\log_{2}\left(x_{min}\right)-\log_{2}\left(\alpha-1\right)+\left(\frac{\alpha}{\alpha-1}\right)$\tabularnewline
\hline 
\end{tabular}

\end{adjustbox}

\protect\caption{Studied PDFs (left column) with their corresponding analytical differential
entropies (right column). \label{tab:PDFsandDiffEntrop}}
\end{table}

\section{Results}

In this section, comparisons of theoretical
vs quantized differential entropy for the PDFs considered are shown. Next,
we provide differential complexity results ($E'_{D}$, $S_{D}$,
and $C_{D}$) for the mentioned PDFs. Furthermore, in the case of power-laws,
we also provide and discuss the corresponding complexity measures
results for real world phenomena, already described in~\cite{Newman2004}.
Also, it is worth noting that, since for quantized $H\left(f\right)$
of the power-law yielded poor results, the power-law's analytical
$H\left(f\right)$ form was used.

\subsection{Theoretical vs Quantized Differential Entropies}

\begin{figure}[H]
\begin{adjustbox}{width=1\textwidth,center}

\includegraphics{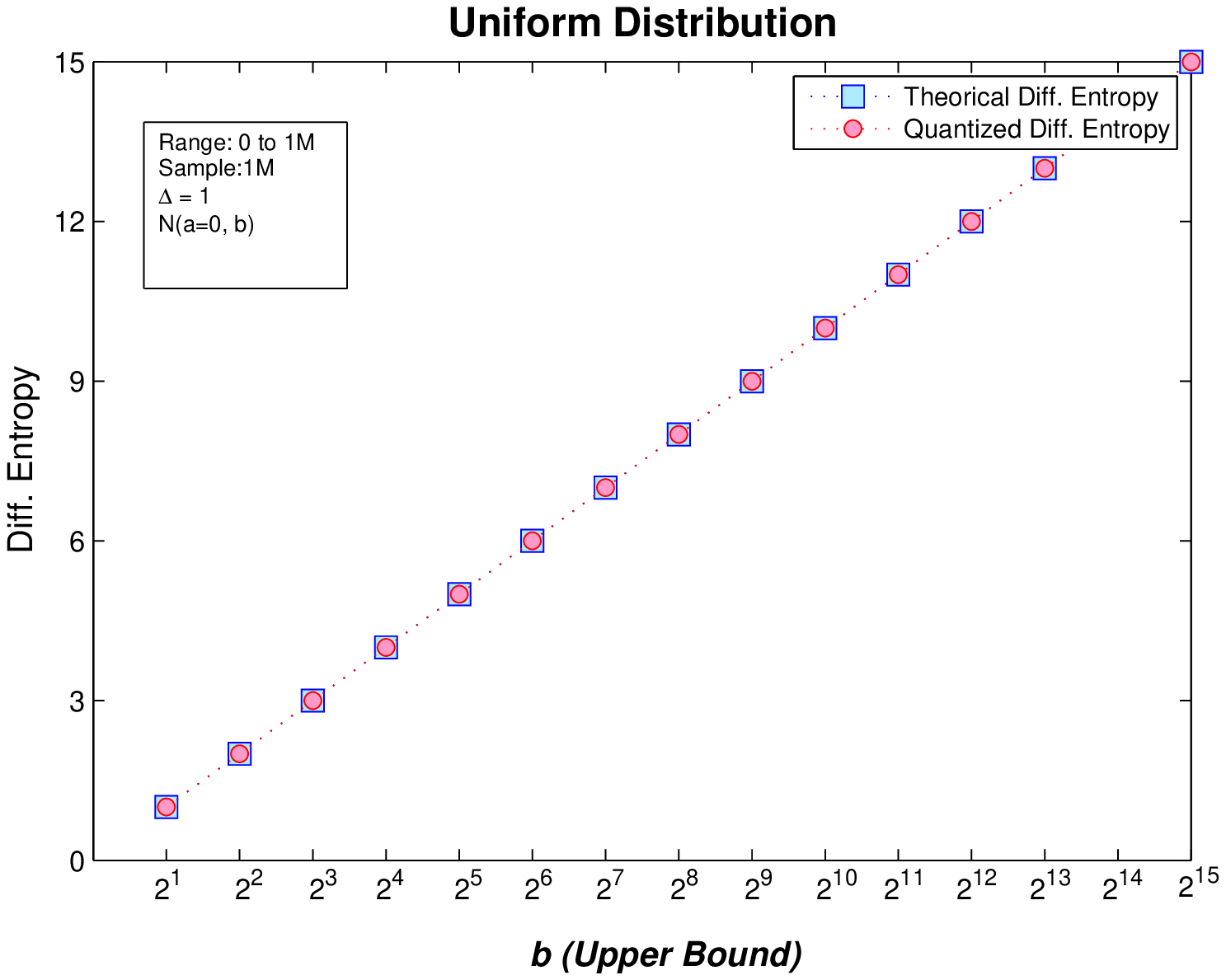}\hfill{}\includegraphics{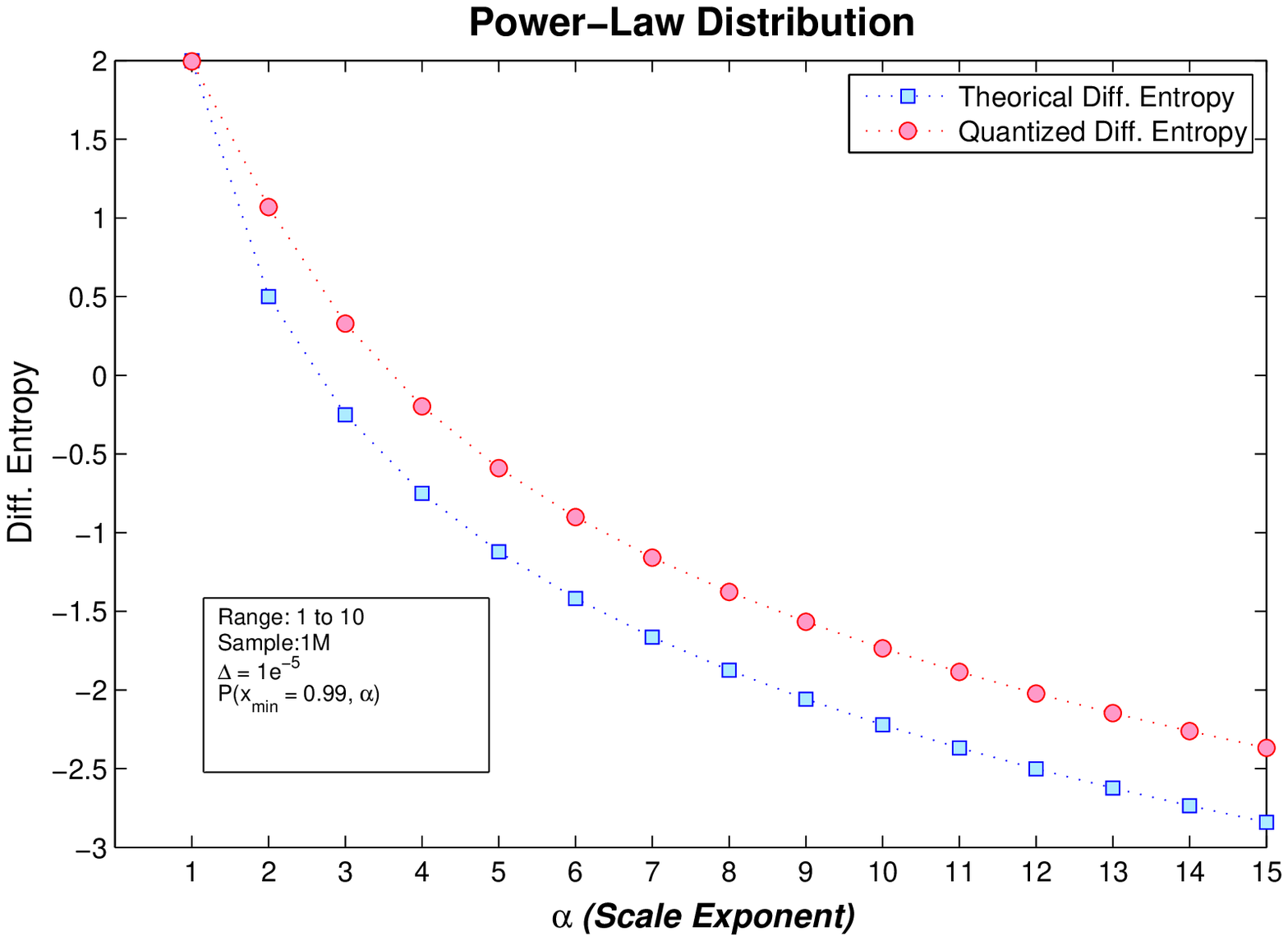}

\end{adjustbox}

\protect\caption{Theoretical and quantized differential entropies for the uniform and power-law
distributions.\label{fig:Differential-Entropy-forUnifAndPowLaw}}
\end{figure}

Numerical results of theoretical and quantized differential entropies are shown in Figs.~\ref{fig:Differential-Entropy-forUnifAndPowLaw}
and~\ref{fig:Differential-Entropy-forGaussianDist}. Analytical $H\left(f\right)$
results are displayed in blue, whereas the quantized $H\left(X^{\Delta}\right)'$
ones are shown in red. For each PDF, a sample of one million
(\emph{i.e.} $1\times10^{6}\equiv\text{1M}$) points where employed for calculations.
The bin size $\Delta$ required by $H\left(X^{\Delta}\right)'$,
is obtained as the ratio $\Delta=\frac{Range}{\left|Sample\right|}$.
However, the value of $\Delta$ has considerable influence in the resulting
quantized differential entropy.

\subsubsection{Uniform Distribution.}

The results for U were expectable. We tested
several values of the cardinality of $P\left(X\right)$, such that
$b=2^{i}\mid i=1,\ldots,15$. Using the analytical $H\left(f\right)$
formula of Table~\ref{tab:PDFsandDiffEntrop}, the quantized $H\left(X^{\Delta}\right)'$,
and $\Delta=1$ we achieved exactly the same differential entropy
values. Results for U are shown in the left side of Fig.~\ref{fig:Differential-Entropy-forUnifAndPowLaw}.
As was mentioned earlier, as the cardinality of the distribution grows,
so does the differential entropy of U. 

\subsubsection{Normal Distribution.}

Results for the Gaussian distribution were less trivial. As in the
U case, we calculate both $H\left(f\right)$ and $H\left(X^{\Delta}\right)'$,
for a fixed $\mu=0$, and modified the standard deviation parameter
such that, $\sigma=2^{i}\mid i=0,1,\ldots,14$. Notice that the first tested distribution is the
\emph{standard normal distribution}.

\begin{figure}[H]
\begin{adjustbox}{width=1\textwidth,center}
\includegraphics{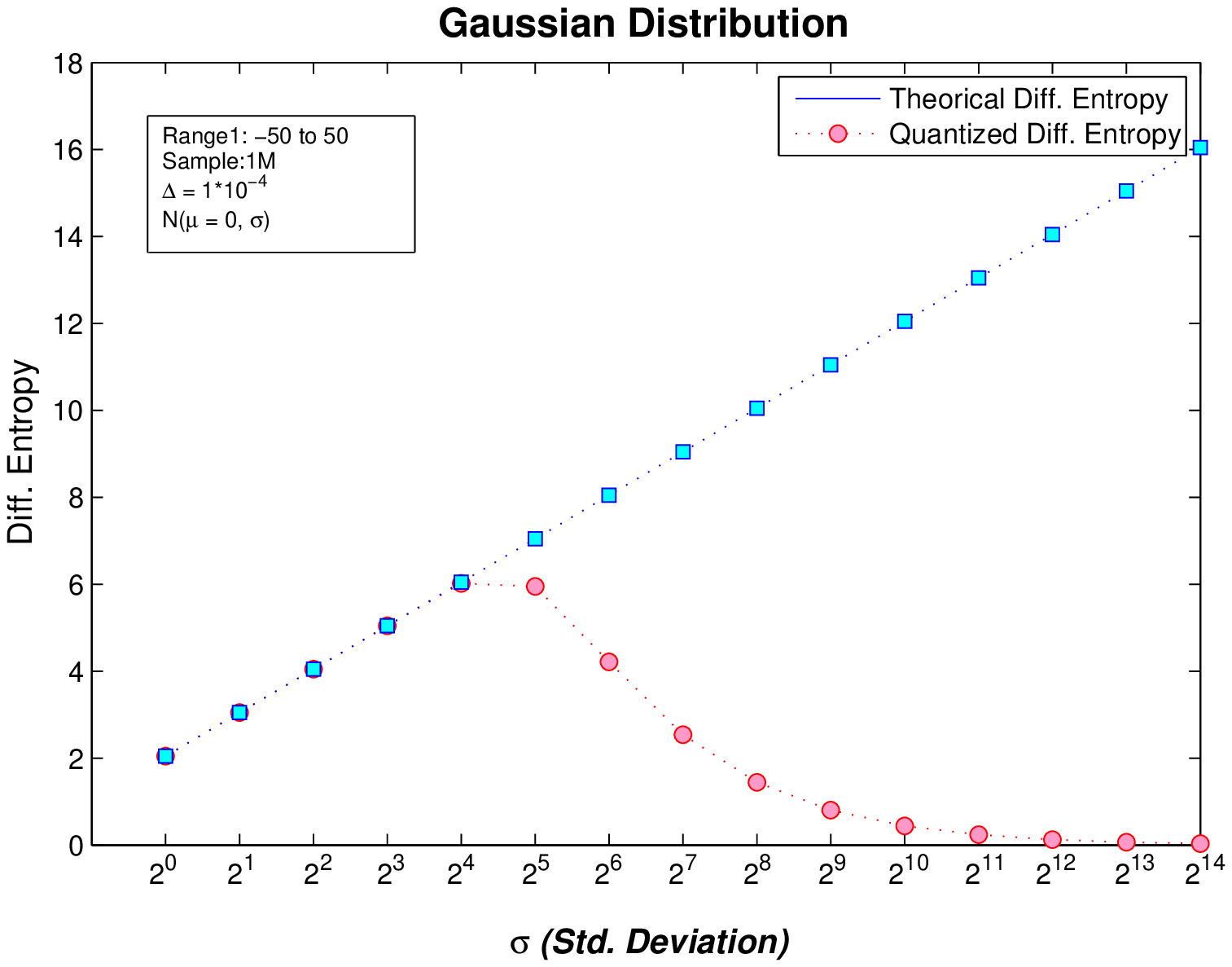}\hfill{}\includegraphics{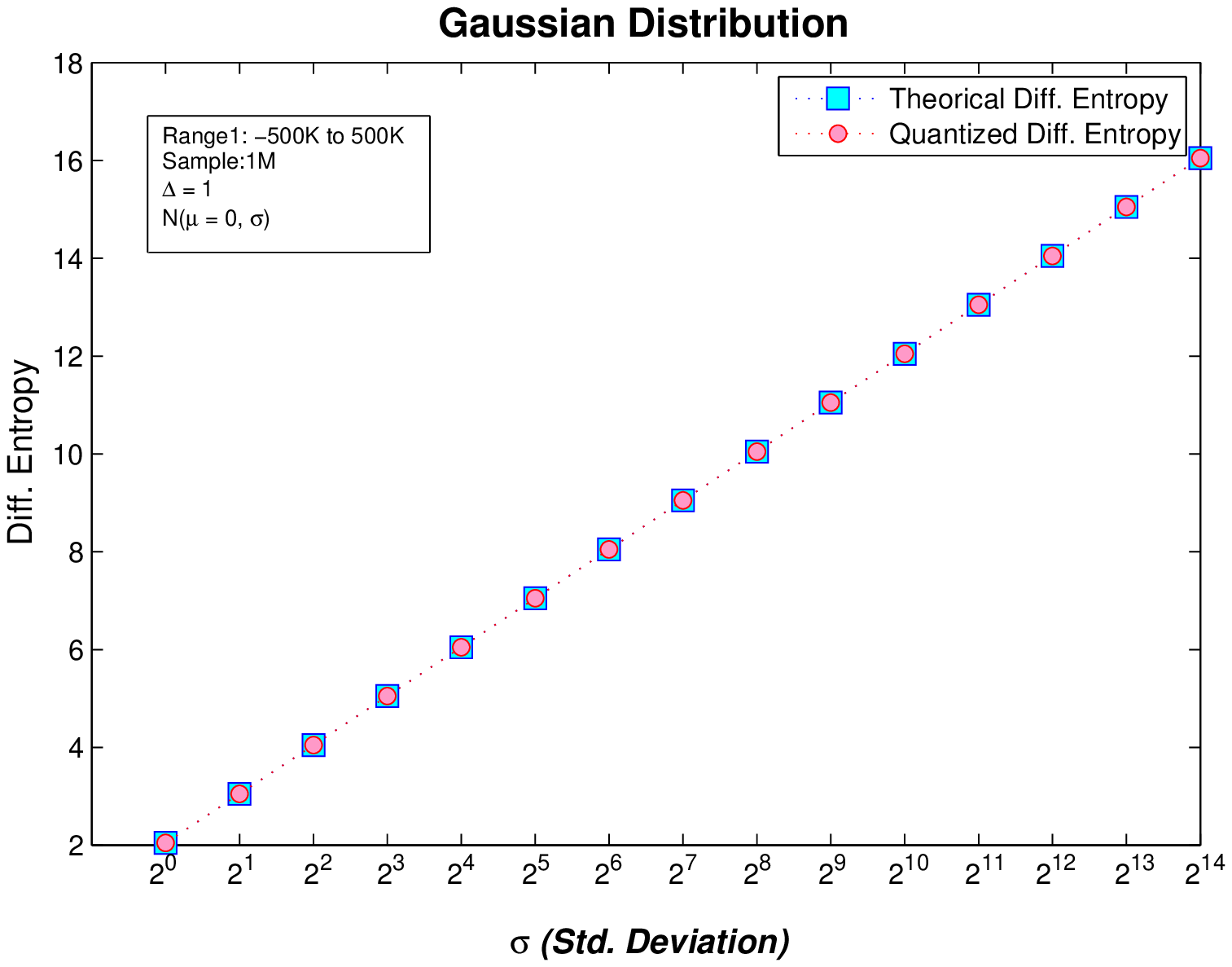}

\end{adjustbox}

\protect\caption{Two comparisons of theoretical vs quantized differential entropy for
the Gaussian distribution.\label{fig:Differential-Entropy-forGaussianDist}}

\end{figure}

In Fig.~\ref{fig:Differential-Entropy-forGaussianDist}, results obtained
for the n-bit quantized differential entropy, and for the analytical
form of Table~\ref{tab:PDFsandDiffEntrop} are shown. Moreover, we
displayed two cases of the normal distribution: the left side of Fig.
\ref{fig:Differential-Entropy-forGaussianDist} shows results for
$P\left(X\right)$ with range $\left[-50,50\right]$ and a bin size,
$\Delta=\frac{100}{1M}=1\times10^{-4}$,whereas, right side provides
results for a $P\left(X\right)$ with range $\left[-500e3,500e3\right]$
and $\Delta=1$. It is worth noting that, in the former case the quantized
differential entropy shows a discrepancy with $H\left(f\right)$ after
only $\sigma=2^{4}=16$, which quickly increases with growing $\sigma$.
On the other hand, for the latter case there is an almost perfect
match between the analytical and quantized differential entropies,
however, the same mismatch will be observed if the standard deviation
parameter is allowed to grow unboundedly $(\sigma\rightarrow\infty)$.
Nonetheless, this is a consequence of how $H\left(X^{\Delta}\right)'$
is computed. As mentioned earlier, as $\Delta\rightarrow0$ the value
of each quantized $X^{\Delta}$ grows towards $-\infty$. Therefore,
in the G case, it seems convenient employing a Probability Mass Function
(PMF) rather than a PDF. Consequently, the experimental setup of right
side image of Fig.~\ref{fig:Differential-Entropy-forGaussianDist}
is employed for the calculation of the continuous complexity measures
of G.

\subsubsection{Power-Law Distribution.}

Results for the power-law distribution are shown in the right side of Fig.~\ref{fig:Differential-Entropy-forUnifAndPowLaw}. In both U and G, a PMF instead of
a PDF was used to avoid cumbersome results (as depicted in the corresponding
images). However, for the power-law distribution, the use of a PDF is rather convenient.
As shown in Fig.~\ref{fig:Cmplxty-Msrs-PL} and highlighted
by~\cite{Clauset2009}, $x_{min}$ has a considerable impact on the value of
$H\left(f\right)$. For Fig.~\ref{fig:Differential-Entropy-forUnifAndPowLaw},
the range employed was $\left[1,50\right]$, with a bin size of $\Delta=1\times10^{-5}$,
a $x_{min}=0.99$, and modified the scale exponent parameter such
that, $\alpha=i\mid i=1,\ldots,15$. For this particular setup, we
can observe that as $\alpha$ increases, $H\left(f\right)$ and $H\left(X^{\Delta}\right)'$
decreases its value towards $-\infty$. This effect is consequence
of increasing the scale of the Power-law such that, the slope of the
function in a log-log space, approaches to zero. In this sense, with
larger $\alpha$'s, the $P\left(X\right)$ becomes closer to a Dirac delta
distribution, thus, $H\left(f\right)\rightarrow-\infty$. However,
as will be discussed later, for larger $\alpha$'s larger $x_{min}$
values are required, in order for $H\left(f\right)$ to display positive
values.

\subsection{Differential Complexity: $E{}_{D}$, $S_{D}$, and $C_{D}$}

U results are trivial: $E{}_{D}=1$,
and $S_{D},C_{D}=0$. For each upper bound of U, $E'_{D}=\frac{H\left(U\right)}{H\left(U\right)}=1$,
which is exactly the same as its discrete counterpart. Thus, U results
are not considered in the following analysis. 

Continuous complexity results for G and PL are
shown in Figs.~\ref{fig:Cmplxty-Msrs-Gauss} and~\ref{fig:Cmplxty-Msrs-PL},
respectively. In the following we provide details of these measures.

\subsubsection{Normal Distribution.}

It was stated in Section~\ref{sub:Enter-Differential-Complexity}
that, the size of the alphabet is given by the function $ind\left(P\left(X\right)\right)$.
This rule establishes a valid cardinality such that $P\left(X\right)>0$,
thus, only those states with a positive probability are considered. For
$P\left(X^{\Delta}\right)$, such operation can be performed. Nevertheless,
when the analytical $H\left(f\right)$ is used, the proper cardinality
of the set is unavailable. Therefore, in the Gaussian distribution
case, we tested two criteria for selecting the value of \emph{b}: 
\begin{enumerate}
\item $\sum_{x_{i}}ind\left(\cdot\right)$ is employed for $H\left(X^{\Delta}\right)'$ 
\item A constant with a large value ($C=1\times10^{6}$)
is used for the analytical formula of $H\left(f\right)$. 
\end{enumerate}

\begin{figure}[H]
\begin{adjustbox}{width=1\textwidth,center}

\includegraphics{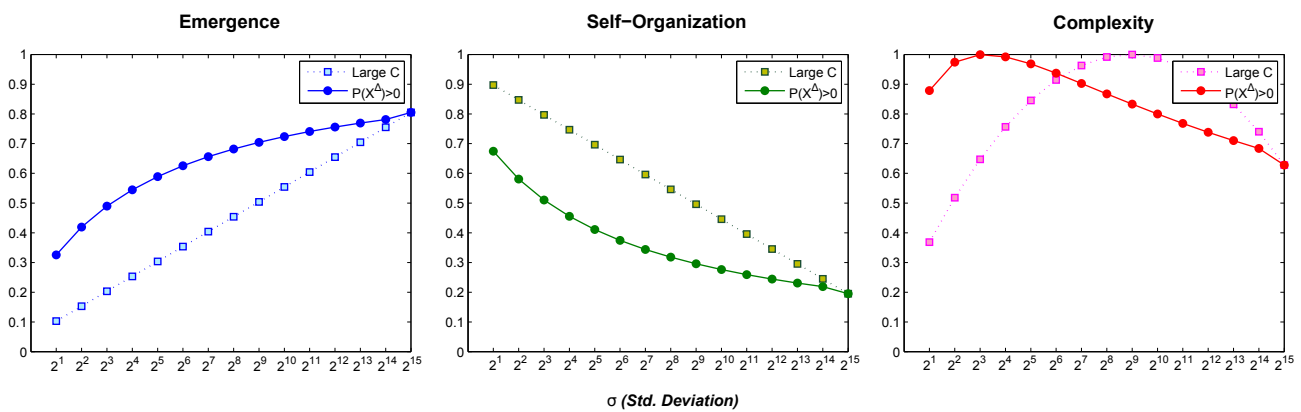}

\end{adjustbox}

\protect\caption{Complexity of the Gaussian distribution.\label{fig:Cmplxty-Msrs-Gauss}}

\end{figure}

In Fig.~\ref{fig:Cmplxty-Msrs-Gauss}, solid dots are used
when \emph{K} is equal to the cardinality of $P\left(X\right)>0$,
whereas solid squares are used for an arbitrary large constant.
Moreover, for the quantized case of $P\left(G\right)$ , Table~\ref{tab:bAndKforGaussDistr}
shows the cardinality for each sigma, $b'_{i}$, and its corresponding
\emph{$K_{i}$}. As it can be observed, for a large normalizing constant
\emph{K}, a logarithmic relation is displayed for $E{}_{D}$ and $S_{D}$.
Also, the maximum $C_{D}$ is achieved for $\sigma=2^{8}=256$, which
is where $E{}_{D}=S_{D}$. However, for $H\left(X^{\Delta}\right)'$
the the maximum $C_{D}$ is found around
$\sigma=2^{1,2,3}=2,4,8$, such that $C_{D}\leq\epsilon\mid\epsilon\rightarrow0$.
A word of advise must be made here. The required cardinality to normalize
the continuous complexity measures such that $0\leq E{}_{D},S_{D},C_{D}\leq1$,
must have a lower bound. This bound should be related to the scale
of the $P\left(X\right)$~\cite{Landsberg1994}, and the quantization
size $\Delta$. In our case, when a large cardinality $\left|U\right|=1\times10^{6}$,
and $\Delta=1$ are used, the normalizing constant \emph{flattens}
$E_{D}$ results respect those obtained by $b'$; moreover, the large
constant increases $S_{D}$, and takes greater standard deviations
for achieving the maximum $C_{D}$. However, these complexity results
are rather artificial in the sense that, if we arbitrarly let $\left|U\right|\rightarrow\infty$
then trivially we will obtain $E{}_{D}=0,\, S_{D}=1,\text{ and }C_{D}=0$.
Moreover, it has been stated that the cardinality of $P\left(X\right)$
should be employed as a proper size of b~\cite{SinghEntrop2013}.
Therefore, when $H\left(X^{\Delta}\right)'$ is employed, the cardinality
of $P\left(X\right)>0$ must be used. On the contrary, when $H\left(f\right)$
is employed, a coarse search for increasing alphabet sizes could be
used so that the maximal $H\left(f\right)$ satisfies $\frac{H\left(f\right)}{H\left(U\right)}\leq1$.

\begin{table}
\begin{centering}
\begin{tabular}{|c|c|c|c|}
\hline 
$\sigma$ & $b'=\sum ind\left(Pr\left(X\right)>0\right)$ & $H\left(U\right)=\log_{2}\left(b'\right)$ & $K=\frac{1}{H\left(U\right)}$\tabularnewline
\hline 
\hline 
$2^{0}=1$ & 78 & 6.28 & 0.16\tabularnewline
\hline 
$2^{1}=2$ & 154 & 7.26 & 0.14\tabularnewline
\hline 
$2^{2}=4$ & 308 & 8.27 & 0.12\tabularnewline
\hline 
$2^{3}=8$ & 616 & 9.27 & 0.11\tabularnewline
\hline 
$2^{4}=16$ & 1232 & 10.27 & 0.10\tabularnewline
\hline 
$2^{5}=32$ & 2464 & 11.27 & 0.09\tabularnewline
\hline 
$2^{6}=64$ & 4924 & 12.27 & 0.08\tabularnewline
\hline 
$2^{7}=128$ & 9844 & 13.27 & 0.075\tabularnewline
\hline 
$2^{8}=256$ & 19680 & 14.26 & 0.0701\tabularnewline
\hline 
$2^{9}=512$ & 39340 & 15.26 & 0.0655\tabularnewline
\hline 
$2^{10}=1024$ & 78644 & 16.26 & 0.0615\tabularnewline
\hline 
$2^{11}=2048$ & 157212 & 17.26 & 0.058\tabularnewline
\hline 
$2^{12}=4096$ & 314278 & 18.26 & 0.055\tabularnewline
\hline 
$2^{13}=8192$ & 628258 & 19.26 & 0.0520\tabularnewline
\hline 
$2^{14}=16384$ & 1000000 & 19.93 & 0.050\tabularnewline
\hline 
\end{tabular}
\par\end{centering}

\protect\caption{Alphabet size $b'$, and its corresponding normalizing \emph{K} constant
for the normal distribution G.}
\label{tab:bAndKforGaussDistr}

\end{table}

\subsubsection{Power-Law Distribution}

In this case, $H\left(f\right)$ rather than $H\left(X^{\Delta}\right)'$
is used for computational convenience. Although the cardinality of
$P\left(X\right)>0$ is not available, by simply substituting $p\left(x_{i}\right)>0\mid x=\left\{ 1,\ldots,1\times10^{6}\right\} $
we can see that the condition is fulfilled by the whole set. Therefore,
the large $C$ criterium, earlier detailed, is used. Still,
given that a numerical power-law distribution is given by two parameters,
a lower bound $x_{min}$ and the scale exponent $\alpha$, we depict
our results in 3D in Fig.~\ref{fig:Cmplxty-Msrs-PL}. From left to
right, $E{}_{D},\, S_{D},\text{ and }C_{D}$ for the power-law distribution
are shown, respectively. In the three images, the same coding is used:
x-axis displays the scale exponent ($\alpha$) values, y-axis shows
$x_{min}$ values, and z-axis depicts the continuous measure
values; lower values of $\alpha$ are displayed in dark blue, turning
into reddish colors for larger exponents.

As it can be appreciated in Fig.~\ref{fig:Cmplxty-Msrs-PL}, for small
$x_{min}$ (\emph{e.g.} $x_{min}=1$) values, low emergence is produced despite
the scale exponent. Moreover, maximal self-organization ($\text{\emph{i.e.} }S_{D}=1$)
is quickly achieved (\emph{i.e.} $\alpha=4$), providing a PL with at most
fair complexity values. However, if we let $x_{min}$ take larger numbers,
$E_{D}$ grows, achieving the maximal complexity (\emph{i.e.} $C_{D}\approx0.8$)
of this experimental setup at $x_{min}=15,\,\alpha=1$. This behavior
is also observed for other scale exponent values, where emergence
of new information is produced as the $x_{min}$ value grows. Furthermore,
it has been stated that for $P\left(X\right)$ displays a power-law
behavior it is required that $\forall x_{i}\in P\left(X\right)\mid x_{i}>x_{min}$
\cite{Virkar2012}. Thus, for every $\alpha$ there should be an $x_{min}$
such that $E_{D}>0$. Moreover, for larger scale exponents, larger
$x_{min}$ values are required for the distribution shows emergence
of new information at all.

\subsection{Real World Phenomena and their Complexity}

Data of phenomena that follows a power law is provided in Table~\ref{tab:PowLawParamAndComplexityRealWorldPhenom}.
These power-laws have been studied by~\cite{Newman2004,Clauset2009,Virkar2012}, and the power-law parameters were published by~\cite{Newman2004}.
The phenomena in the table mentioned above compromises data from: 
\begin{enumerate}
\item Numbers of occurrences of words in the novel Moby Dick by Hermann
Melville.
\item Numbers of citations to scientific papers published in 1981, from
the time of publication until June 1997.
\item Numbers of hits on websites by users of America Online Internet services
during a single day.
\item Number of received calls to A.T.\&T. U.S. long-distance telephone
services on a single day.
\item Earthquake magnitudes occurred in California between 1910 and 1992.
\item Distribution of the diameter of moon craters.
\item Peak gamma-ray intensity of solar flares between 1980 and 1989.
\item War intensity between 1816\textendash 1980, where intensity is a
formula related to the number of deaths and warring nations populations.
\item Frequency of family names accordance with U.S. 1990 census.
\item Population per city in the U.S. in agreement with U.S. 2000 census.
\end{enumerate}

More details about these power-laws can be found in~\cite{Newman2004,Clauset2009,Virkar2012}.

\begin{figure}[H]
\begin{adjustbox}{width=1\textwidth,center}

\includegraphics{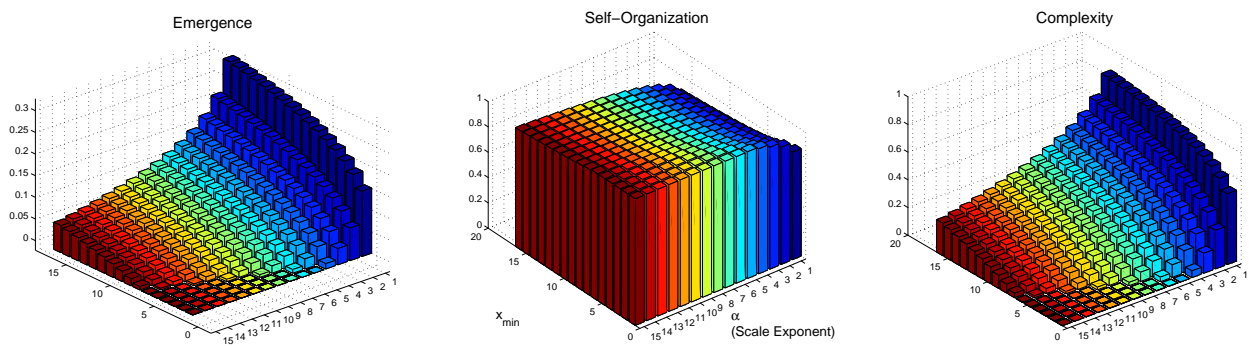}

\end{adjustbox}

\protect\caption{Complexity measures for the Power-Law. Lower values of the scale exponent $\alpha$ are displayed in dark blue, colors turns into reddish for larger scale exponents. \label{fig:Cmplxty-Msrs-PL}}
\end{figure}

For each phenomenon, the corresponding differential entropy and complexity
measures are shown in Table~\ref{tab:PowLawParamAndComplexityRealWorldPhenom}. Furthermore, we also provide 
Table~\ref{tab:ESC_ColorCoding} which is a color coding for complexity measures proposed in ~\cite{Fernandez2013Information-Mea}. Five colors are employed to simplify the different value ranges of $E_D$, $S_D$, and $C_D$ results.
According to the nomenclature suggested in~\cite{Fernandez2013Information-Mea},
results for these sets show that, very high complexity $0.8\leq C_{D}\leq1$ 
is obtained by the number of citations set (\emph{i.e.} 2), and intensity
of solar flares (\emph{i.e.} 7). High complexity, $0.6\leq C_{D}<0.8$ is
obtained for received telephone calls (\emph{i.e.} 4), intensity of wars
(\emph{i.e.} 8), and frequency of family names (\emph{i.e.} 9). Fair complexity
$0.4\leq C_{D}<0.6$ is displayed by earthquakes magnitude (\emph{i.e.} 5),
and population of U.S. cities (\emph{i.e.} 10). Low complexity, $0.2\leq C_{D}<0.4$
is obtained for frequency of used words in Moby Dick (\emph{i.e.} 1) and web
hits (\emph{i.e.} 3), whereas, moon craters (\emph{i.e.} 6) have very low complexity
$0\leq C_{D}<0.2$. In fact, earthquakes, and web hits, have been
found not to follow a power law~\cite{Clauset2009}. Furthermore,
if such sets were to follow a power-law, a greater value of $x_{min}$
would be required as can be observed in Fig.~\ref{fig:Cmplxty-Msrs-PL}.
In fact, the former case is found for the frequency of words used
in Moby Dick. In~\cite{Newman2004}, parameters of Table~\ref{tab:PowLawParamAndComplexityRealWorldPhenom}
are proposed. However, in~\cite{Clauset2009}, another set of parameters
are estimated (\emph{i.e.} $x_{min}=7,\,\alpha=1.95$). For the more recent
estimated set of parameters, a high complexity is achieved (\emph{i.e.} $C_{D}=0.74$),
which is more consistent with literature about Zipf's law~\cite{Newman2004}.
Lastly, in the case of moon craters, the $x_{min}=0.01$ is rather
a poor choice according to Fig.~\ref{fig:Cmplxty-Msrs-PL}. For
the chosen scale exponent, it would require at least a $x_{min}\approx1$,
for the power-law to produce any information at all. It should be noted that $x_{min}$ can be adjusted to change the values of all measures. Also, it is worth mentioning that if we were to normalize and discretize a power law distribution to calculate its discrete entropy (as in~\cite{Fernandez2013Information-Mea}), all power law distributions present a very high complexity, independently of $x_{min}$ and $\alpha$, precisely because these are normalized. Still, this is not useful for comparing different power law distributions.

\begin{table}[H]
\begin{adjustbox}{width=1\textwidth,center}

\begin{tabular}{|c|c|c|c|c|c|c|c|}
\hline 
 & Phenomenon & $x_{min}$ & $\alpha$ (Scale Exponent) & $H\left(f\right)$ & $E'_{D}$ & $S_{D}$ & $C_{D}$\tabularnewline
\hline 
\hline 
1 & Frequency of use of words & 1 & 2.2 & 1.57 & \cellcolor{red!70} 0.078 & \cellcolor{blue!70}0.92 & \cellcolor{orange!70}0.29\tabularnewline
\hline 
2 & Number of citations to papers & 100 & 3.04 & 7.1 & \cellcolor{orange!70} 0.36 & \cellcolor{green!70}0.64 & \cellcolor{blue!70}0.91\tabularnewline
\hline 
3 & Number of hits on web sites & 1 & 2.4 & 1.23 &\cellcolor{red!70} 0.06 &\cellcolor{blue!70} 0.94 & \cellcolor{orange!70}0.23\tabularnewline
\hline 
4 & Telephone calls received & 10 & 2.22 & 4.85 & \cellcolor{orange!70} 0.24 & \cellcolor{green!70} 0.76 & \cellcolor{green!70} 0.74\tabularnewline
\hline 
5 & Magnitude of earthquakes & 3.8 & 3.04 & 2.38 & \cellcolor{red!70} 0.12 & \cellcolor{blue!70} 0.88 & \cellcolor{yellow!70} 0.42\tabularnewline
\hline 
6 & Diameter of moon craters & 0.01 & 3.14 & -6.27 & \cellcolor{red!70}0  & \cellcolor{blue!70}1 &\cellcolor{red!70} 0 \tabularnewline
\hline 
7 & Intensity of solar flares & 200 & 1.83 & 10.11 & \cellcolor{yellow!70}0.51 &\cellcolor{yellow!70} 0.49 & \cellcolor{blue!70}0.99\tabularnewline
\hline 
8 & Intensity of wars & 3 & 1.80 & 4.15 & \cellcolor{orange!70}0.21 & \cellcolor{green!70}0.79 & \cellcolor{green!70}0.66\tabularnewline
\hline 
9 & Frequency of family names & 10000 & 1.94 & 15.44 & \cellcolor{green!70}0.78 & \cellcolor{orange!70}0.22 & \cellcolor{green!70}0.7\tabularnewline
\hline 
10 & Population of U.S. cities & 40000 & 2.30 & 16.67 & \cellcolor{blue!70}0.83 & \cellcolor{red!70}0.17 & \cellcolor{yellow!70}0.55\tabularnewline
\hline 
\end{tabular}

\end{adjustbox}

\protect\caption{Power-Law parameters and information-based measures of real world
phenomena \label{tab:PowLawParamAndComplexityRealWorldPhenom}}

\end{table}

\begin{table}[htbp]
  \centering
  \caption{Categories for classifying $E$, $S$, and $C$.}
    \begin{tabular}{|c|c|c|c|c|c|}
    \toprule
    \textbf{Category} & Very High & High  & Fair  & Low   & Very Low \\
    \midrule
    \textbf{Range} & $[0.8,1]$ & $[0.6,0.8)$ & $[0.4,0.6)$ & $[0.2,0.4)$ & $[0,0.2)$ \\
    \midrule
    \textbf{Color} & \cellcolor{blue!70} Blue  & \cellcolor{green!70} Green & \cellcolor{yellow!70} Yellow &  \cellcolor{orange!70} Orange & \cellcolor{red!70} Red \\
    \bottomrule
    \end{tabular}%
    \protect\caption{Color coding for $E_D$, $S_D$, and $C_D$ results 	 
    \label{tab:ESC_ColorCoding}}
\end{table}

\section{Discussion}

The relevance of the work presented here lies in the fact that it is now possible to calculate measures of emergence, self-organization, and complexity directly from probability distributions, without needing access to raw data. Certainly, the \emph{interpretation} of the measures is not given, as this will depend on the use we make of the measures for specific purposes. 

From exploring the parameter space of the uniform, normal, and scale-free distributions, we can corroborate that high complexity values require a form of \emph{balance} between extreme cases. On the one hand, uniform distributions, by definition, are homogeneous and thus all states are equiprobable, yielding the highest emergence. This is also the case of normal distributions with a very large standard deviation and for power law distributions with an exponent close to zero. On the other hand, highly biased distributions (very small standard deviation in G or very large exponent in PL) yield a high self-organization, as few states accumulate most of the probability. Complexity is found between these two extremes. From the values of $\sigma$ and $\alpha$, this coincides with a broad range of phenomena. This does not tell us something new: complexity is common. The relevant aspect is that this provides a common framework to study of the processes that lead phenomena to have a high complexity~\cite{GershensonLenaerts2008}. It should be noted that this also depends on the time scales at which change occurs~\cite{Cocho2015}.

In this context, it is interesting to relate our results with information adaptation~\cite{Haken2015}. In a variety of systems, adaptation takes place by \emph{inflating} or \emph{deflating} information, so that the ``right" balance is achieved. Certainly, this precise balance can change from system to system and from context to context. Still, the capability of information adaptation has to be correlated with complexity, as the measure also reflects a balance between emergence (inflated information) and self-organization (deflated information). 

As a future work, it will be interesting to study the relationship between complexity and semantic information. There seems to be a connection with complexity as well, as we have proposed a measure of \emph{autopoiesis} as the ratio of the complexity of a system over the complexity of its environment~\cite{Fernandez2013Information-Mea,Gershenson2015Requisite-Varie}. These efforts should be valuable in the study of the relationship between information and meaning, in particular in cognitive systems.

Another future line of research  lies in the relationship between the proposed measures and complex networks~\cite{Newman:2003,NewmanEtAl2006,Boccaletti2006cxNets,GershensonProkopenko:2011}, exploring questions such as: how does the topology of a network affect its dynamics? How much can we predict the dynamics of a network based on its topology? What is the relationship between topological complexity and dynamic complexity? How controllable are networks~\cite{Motter2015} depending on their complexity?



\acknowledgments{Acknowledgments}

G.SB. was supported by the Universidad Nacional Autónoma de México (UNAM) under grant CJIC/CTIC/0706/2014.
C.G. was supported by CONACYT projects 212802, 221341, and SNI membership 47907. 


\authorcontributions{Author Contributions}

GSB and CG conceived and designed the experiments, GSB performed the experiments, GSB, NF, and CG wrote the paper.


\conflictofinterests{Conflicts of Interest}

The authors declare no conflict of interest'.


\bibliography{library,bib}

\begin{thebibliography}{-------}
\providecommand{\natexlab}[1]{#1}

\bibitem[Gershenson(2008)]{Cx5Q}
Gershenson, C., Ed.
\newblock {\em Complexity: 5 Questions}; Automatic Peess / VIP,  2008.

\bibitem[Prokopenko \em{et~al.}(2009)Prokopenko, Boschetti, and
  Ryan]{Prokopenko:2008}
Prokopenko, M.; Boschetti, F.; Ryan, A.
\newblock An Information-Theoretic Primer On Complexity, Self-Organisation And
  Emergence.
\newblock {\em Complexity} {\bf 2009}, {\em 15},~11 -- 28.

\bibitem[Gershenson and Fern\'andez(2012)]{GershensonFernandez:2012}
Gershenson, C.; Fern\'andez, N.
\newblock Complexity and Information: Measuring Emergence, Self-organization,
  and Homeostasis at Multiple Scales.
\newblock {\em Complexity} {\bf 2012}, {\em 18},~29--44.

\bibitem[Fern\'andez \em{et~al.}(2014)Fern\'andez, Maldonado, and
  Gershenson]{Fernandez2013Information-Mea}
Fern\'andez, N.; Maldonado, C.; Gershenson, C.
\newblock Information Measures of Complexity, Emergence, Self-organization,
  Homeostasis, and Autopoiesis. In {\em Guided Self-Organization: Inception};
  Prokopenko, M., Ed.; Springer: Berlin Heidelberg,  2014; Vol.~9, {\em
  Emergence, Complexity and Computation}, pp. 19--51.

\bibitem[Shannon(1948)]{Shannon1948}
Shannon, C.E.
\newblock A mathematical theory of communication.
\newblock {\em Bell System Technical Journal} {\bf 1948}, {\em 27},~379--423
  and 623--656.

\bibitem[Gershenson and Heylighen(2003)]{GershensonHeylighen2003a}
Gershenson, C.; Heylighen, F.
\newblock When Can We Call a System Self-Organizing?
\newblock  Advances in Artificial Life, 7th European Conference, {ECAL} 2003
  {LNAI} 2801; Banzhaf, W.; Christaller, T.; Dittrich, P.; Kim, J.T.; Ziegler,
  J., Eds.; Springer: Berlin,  2003; pp. 606--614.

\bibitem[Langton(1990)]{Langton1990}
Langton, C.G.
\newblock Computation at the Edge of Chaos: Phase Transitions and Emergent
  Computation.
\newblock {\em Physica D} {\bf 1990}, {\em 42},~12--37.

\bibitem[Kauffman(1993)]{Kauffman1993}
Kauffman, S.A.
\newblock {\em The Origins of Order}; Oxford University Press: Oxford, UK,
  1993.

\bibitem[Lopez-Ruiz \em{et~al.}(1995)Lopez-Ruiz, Mancini, and
  Calbet]{LopezRuiz:1995}
Lopez-Ruiz, R.; Mancini, H.L.; Calbet, X.
\newblock A statistical measure of complexity.
\newblock {\em Physics Letters A} {\bf 1995}, {\em 209},~321--326.

\bibitem[Zubillaga \em{et~al.}(2014)Zubillaga, Cruz, Aguilar, Zapot\'ecatl,
  Fern\'andez, Aguilar, Rosenblueth, and
  Gershenson]{Zubillaga2014Measuring-the-C}
Zubillaga, D.; Cruz, G.; Aguilar, L.D.; Zapot\'ecatl, J.; Fern\'andez, N.;
  Aguilar, J.; Rosenblueth, D.A.; Gershenson, C.
\newblock Measuring the Complexity of Self-organizing Traffic Lights.
\newblock {\em Entropy} {\bf 2014}, {\em 16},~2384--2407.

\bibitem[Amoretti and Gershenson(2015)]{Amoretti2015Measuring-the-c}
Amoretti, M.; Gershenson, C.
\newblock Measuring the complexity of adaptive peer-to-peer systems.
\newblock {\em Peer-to-Peer Networking and Applications} {\bf 2015}, pp. 1--16.

\bibitem[Febres \em{et~al.}(2015)Febres, Jaffe, and
  Gershenson]{Febres2013Complexity-meas}
Febres, G.; Jaffe, K.; Gershenson, C.
\newblock Complexity measurement of natural and artificial languages.
\newblock {\em Complexity} {\bf 2015}, {\em 20},~25--48.

\bibitem[Santamar\'ia-Bonfil \em{et~al.}(2016)Santamar\'ia-Bonfil,
  Reyes-Ballesteros, and Gershenson]{Santamaria-Bonfil2016Wind-speed-fore}
Santamar\'ia-Bonfil, G.; Reyes-Ballesteros, A.; Gershenson, C.
\newblock Wind speed forecasting for wind farms: A method based on support
  vector regression.
\newblock {\em Renewable Energy} {\bf 2016}, {\em 85},~790--809.

\bibitem[Fern{\'{a}}ndez \em{et~al.}(Submitted)Fern{\'{a}}ndez, Villate,
  Ter{\'{a}}n, Aguilar, and Gershenson]{FernandezCxLakes}
Fern{\'{a}}ndez, N.; Villate, C.; Ter{\'{a}}n, O.; Aguilar, J.; Gershenson, C.
\newblock {Complexity of Lakes in a Latitudinal Gradient}.
\newblock {\em Ecological Complexity} {\bf Submitted}.

\bibitem[Cover and Thomas(2005)]{Cover2005}
Cover, T.; Thomas, J.
\newblock {\em {Elements of Information Theory}}; Wiley,  2005; pp. 1--748,
  \href{http://xxx.lanl.gov/abs/ISBN 0-471-06259-6}{{\normalfont [ISBN
  0-471-06259-6]}}.

\bibitem[Michalowicz \em{et~al.}(2013)Michalowicz, Nichols, and
  Bucholtz]{Michalowicz2013}
Michalowicz, J.; Nichols, J.; Bucholtz, F.
\newblock {\em {Handbook of Differential Entropy}}; Chapman {\&} Hall/CRC,
  2013; pp. 19--43.

\bibitem[Haken and Portugali(2015)]{Haken2015}
Haken, H.; Portugali, J.
\newblock {\em {Information Adaptation: The Interplay Between Shannon
  Information and Semantic Information in Cognition}}; SpringerBriefs in
  Complexity, Springer International Publishing,  2015.

\bibitem[Heylighen and Joslyn(2003)]{Heylighen2001}
Heylighen, F.; Joslyn, C.
\newblock {Cybernetics and Second-Order Cybernetics}. In {\em Encycl. Phys.
  Sci. Technol.}; Elsevier,  2003; Vol.~4, pp. 155--169.

\bibitem[Ashby(1956)]{Ashby1956}
Ashby, W.R.
\newblock {\em An Introduction to Cybernetics}; Chapman \& Hall: London,  1956.

\bibitem[Michalowicz \em{et~al.}(2008)Michalowicz, Nichols, and
  Bucholtz]{Michalowicz2008}
Michalowicz, J.V.; Nichols, J.M.; Bucholtz, F.
\newblock {Calculation of differential entropy for a mixed Gaussian
  distribution}.
\newblock {\em Entropy} {\bf 2008}, {\em 10},~200--206.

\bibitem[Calmet and Calmet(2005)]{Calmet2005}
Calmet, J.; Calmet, X.
\newblock Differential Entropy on Statistical Spaces,
  \href{http://xxx.lanl.gov/abs/0505397}{{\normalfont
  [arXiv:cond-mat/0505397]}}.
\newblock arxiv cond-mat/0505397.

\bibitem[Yeung(2008)]{Yeung2008}
Yeung, R.
\newblock {\em {Information Theory and Network Coding}}, 1 ed.; Springer,
  2008; pp. 229--256.

\bibitem[Singh(2013)]{SinghEntrop2013}
Singh, V.
\newblock {\em {Entropy Theory and its Application in Environmental and Water
  Engineering}}; John Wiley {\&} Sons, Ltd: Chichester, UK,  2013; pp. 1--136.

\bibitem[Gershenson(2012)]{Gershenson2012}
Gershenson, C.
\newblock {The Implications of Interactions for Science and Philosophy}.
\newblock {\em Found. Sci.} {\bf 2012}, {\em 18},~781--790.

\bibitem[Sharma and Sharma(2006)]{Sharma2006}
Sharma, K.; Sharma, S.
\newblock {Power Law and Tsallis Entropy: Network Traffic and Applications}. In
  {\em Chaos, Nonlinearity, Complex.}; Springer Berlin Heidelberg: Berlin,
  Heidelberg,  2006; Vol. 178, pp. 162--178.

\bibitem[Dover(2004)]{Dover2004}
Dover, Y.
\newblock {A short account of a connection of power laws to the information
  entropy}.
\newblock {\em Phys. A Stat. Mech. its Appl.} {\bf 2004}, {\em 334},~591--599,
  \href{http://xxx.lanl.gov/abs/0309383}{{\normalfont
  [arXiv:cond-mat/0309383]}}.

\bibitem[Bashkirov and Vityazev(2000)]{Bashkirov2000}
Bashkirov, A.; Vityazev, A.
\newblock {Information entropy and power-law distributions for chaotic
  systems}.
\newblock {\em Phys. A Stat. Mech. its Appl.} {\bf 2000}, {\em 277},~136--145.

\bibitem[Ahsanullah \em{et~al.}(2014)Ahsanullah, Kibria, and
  Shakil]{Ahsanullah2014}
Ahsanullah, M.; Kibria, B.; Shakil, M.
\newblock {\em {Normal and Student's t-Distributions and Their Applications}};
  Vol.~4, {\em Atlantis Studies in Probability and Statistics}, Atlantis Press:
  Paris,  2014.

\bibitem[Box \em{et~al.}(2008)Box, Jenkins, and Reinsel]{Box-Jenkins2008}
Box, G.; Jenkins, G.; Reinsel, G.
\newblock {\em {Time Series Analysis: Forecasting and Control}}, 4th ed. ed.;
  Wiley,  2008.

\bibitem[Mitzenmacher(2009)]{Mitzenmacher}
Mitzenmacher, M.
\newblock {A Brief History of Generative Models for Power Law and Lognormal
  Distributions},  2009,  \href{http://xxx.lanl.gov/abs/0402594v3}{{\normalfont
  [arXiv:arXiv:cond-mat/0402594v3]}}.

\bibitem[Mitzenmacher(2001)]{Mitzenmacher2001}
Mitzenmacher, M.
\newblock {A brief history of generative models for power law and lognormal
  distributions}.
\newblock {\em Internet Math.} {\bf 2001}, {\em 1},~226 -- 251.

\bibitem[Clauset \em{et~al.}(2009)Clauset, Shalizi, and Newman]{Clauset2009}
Clauset, A.; Shalizi, C.R.; Newman, M.E.J.
\newblock {Power-Law Distributions in Empirical Data}.
\newblock {\em SIAM Rev.} {\bf 2009}, {\em 51},~661--703.

\bibitem[Frigg and Werndl(2011)]{Frigg2011}
Frigg, R.; Werndl, C.
\newblock Entropy: A Guide for the Perplexed. In {\em Probabilities in
  physics}; Beisbart, C.; Hartmann, S., Eds.; Oxford University Press: Oxford,
  UK,  2011; pp. 115--142.

\bibitem[Virkar and Clauset(2014)]{Virkar2012}
Virkar, Y.; Clauset, A.
\newblock {Power-law distributions in binned empirical data}.
\newblock {\em Annals of Applied Statistics} {\bf 2014}, {\em 8},~89 -- 119,
  \href{http://xxx.lanl.gov/abs/arXiv:1208.3524v1}{{\normalfont
  [arXiv:1208.3524v1]}}.

\bibitem[Yapage(2014)]{Yapage2014}
Yapage, N.
\newblock {Some Information measures of Power-law Distributions Some
  Information measures of Power-law Distributions}.
\newblock  1st Ruhuna Int. Sci. Technol. Conf.,  2014, pp. 1--6.

\bibitem[Newman(2005)]{Newman2004}
Newman, M.
\newblock {Power laws, Pareto distributions and Zipf's law}.
\newblock {\em Contemp. Phys.} {\bf 2005}, {\em 46},~323--351,
  \href{http://xxx.lanl.gov/abs/0412004}{{\normalfont
  [arXiv:cond-mat/0412004]}}.

\bibitem[Landsberg(1994)]{Landsberg1994}
Landsberg, P.
\newblock {Self-Organization, Entropy and Order}. In {\em On
  Self-Organization}; Mishra, R.K.; Maa{\ss}, D.; Zwierlein, E., Eds.; Springer
  Berlin Heidelberg: Berlin, Heidelberg,  1994; Vol.~61, pp. 157--184.

\bibitem[Gershenson and Lenaerts(2008)]{GershensonLenaerts2008}
Gershenson, C.; Lenaerts, T.
\newblock Evolution of Complexity.
\newblock {\em Artificial Life} {\bf 2008}, {\em 14},~1--3.
\newblock Special Issue on the Evolution of Complexity.

\bibitem[Cocho \em{et~al.}(2015)Cocho, Flores, Gershenson, Pineda, and
  S{\'a}nchez]{Cocho2015}
Cocho, G.; Flores, J.; Gershenson, C.; Pineda, C.; S{\'a}nchez, S.
\newblock Rank Diversity of Languages: Generic Behavior in Computational
  Linguistics.
\newblock {\em PLoS ONE} {\bf 2015}, {\em 10},~e0121898.

\bibitem[Gershenson(2015)]{Gershenson2015Requisite-Varie}
Gershenson, C.
\newblock Requisite Variety, Autopoiesis, and Self-organization.
\newblock {\em Kybernetes} {\bf 2015}, {\em 44},~866--873.

\bibitem[Newman(2003)]{Newman:2003}
Newman, M.E.J.
\newblock The structure and function of complex networks.
\newblock {\em SIAM Review} {\bf 2003}, {\em 45},~167--256.

\bibitem[Newman \em{et~al.}(2006)Newman, Barab{\'a}si, and
  Watts]{NewmanEtAl2006}
Newman, M.; Barab{\'a}si, A.L.; Watts, D.J., Eds.
\newblock {\em The Structure and Dynamics of Networks}; Princeton Studies in
  Complexity, Princeton University Press: Princeton, NJ, USA,  2006.

\bibitem[Boccaletti \em{et~al.}(2006)Boccaletti, Latora, Moreno, Chavez, and
  Hwang]{Boccaletti2006cxNets}
Boccaletti, S.; Latora, V.; Moreno, Y.; Chavez, M.; Hwang, D.U.
\newblock Complex networks: Structure and dynamics.
\newblock {\em Physics Reports} {\bf 2006}, {\em 424},~175 -- 308.

\bibitem[Gershenson and Prokopenko(2011)]{GershensonProkopenko:2011}
Gershenson, C.; Prokopenko, M.
\newblock Complex Networks.
\newblock {\em Artificial Life} {\bf 2011}, {\em 17},~259--261.

\bibitem[Motter(2015)]{Motter2015}
Motter, A.E.
\newblock Networkcontrology.
\newblock {\em Chaos} {\bf 2015}, {\em 25},~--.

\end{thebibliography}
\bibliographystyle{mdpi}


%

\end{document}